\documentclass[9pt,twocolumn,twoside,superscriptaddress]{revtex4-1}

\usepackage{dcolumn} 
\usepackage{bm,amsmath,graphicx} 
\newcommand{\rmg}{\mathrm{g}} 
 
\newcommand{\rmi}{\mathrm{i}}

\newcommand{\rmm}{\mathrm{m}} 
\newcommand{\rmn}{\mathrm{n}}

\newcommand{\rms}{\mathrm{s}}

\newcommand{\rmA}{\mathrm{A}}

\newcommand{\rmM}{\mathrm{M}}

\newcommand{\rmP}{\mathrm{P}}

\newcommand{\rmS}{\mathrm{S}}

\newcommand{\rmW}{\mathrm{W}}

\newcommand{\bfk}{\mathbf{k}}

\newcommand{\bfq}{\mathbf{q}}

 \newcommand{\chalc}{As$_2$S$_3$ }  
  
 \date{\today}

\begin{document}
 \title{ Stimulated Brillouin Scattering in layered media:  nanoscale enhancement of silicon  }

\author{M. J. A. Smith}
\email{michael.j.smith@sydney.edu.au}
\affiliation{Institute of Photonics and Optical Science (IPOS), School of Physics, The University of Sydney, NSW 2006, Australia}
\affiliation{School of Mathematical and Physical Sciences, University of Technology Sydney, NSW 2007, Australia}
\author{C. Wolff}
\affiliation{Center for Nano Optics, University of Southern Denmark, Campusvej 55, DK-5230 Odense M, Denmark}
\author{C. G. Poulton}
\affiliation{School of Mathematical and Physical Sciences, University of Technology Sydney, NSW 2007, Australia}
\author{C. M. de Sterke}
\affiliation{Institute of Photonics and Optical Science (IPOS), School of Physics, The University of Sydney, NSW 2006, Australia}

\begin{abstract}
We report a theoretical study of Stimulated Brillouin Scattering (SBS) in general anisotropic media,
 incorporating the effects of both acoustic   strain and local rotation    in all calculations. We apply our general theoretical framework  to compute the  SBS gain for  layered media with   periodic length scales   smaller  than  all optical and acoustic wavelengths, where such composites   behave like homogeneous anisotropic media. We theoretically predict that a layered medium comprising nanometre-thin layers of silicon and As$_2$S$_3$ glass possesses a bulk SBS gain of $1.28 \times 10^{-9}  \, \rmW^{-1} \, \rmm $. This is more than 500 times larger than the gain coefficient of silicon, and substantially larger than the gain of As$_2$S$_3$. The enhancement is due to a combination of roto-optic, photoelastic, and artificial photoelastic contributions in the composite structure.

\end{abstract}

\maketitle

 Interactions between photons and phonons represent  an important avenue of   research in contemporary photonics and optomechanics \cite{van2015interaction,florez2016brillouin,otterstrom2018silicon}, not only for   the  transmission of light in optical fibres \cite{boyd2003nonlinear,powers2011fundamentals}, but also for   the  design of efficient, small-scale optical devices   \cite{eggleton2013inducing}. One of the most important effects for driving these interactions is  Stimulated Brillouin Scattering (SBS) \cite{powers2011fundamentals}.  In bulk materials, SBS is frequently   described as the resonant excitation of the first acoustic pressure mode of the medium by the optical pump field; the pressure wave acts as a travelling diffraction grating  which scatters the pump and induces a Doppler-shifted returning (Stokes) wave. A significant   issue with this description is that it  only   holds in optically isotropic media where bulk SBS interactions are  only possible with longitudinal acoustic   waves and not with shear acoustic  waves.  In materials possessing   optical  anisotropy, shear acoustic waves and mixed-polarised acoustic waves are   also  SBS-active   \cite{nelson1970measurement,auld1973acoustic,nelson1977dispersive}, due to reduced symmetry   constraints \cite{wolff2014formal}.  Here  we   consider SBS more generally than is frequently described, defining the process as the  inelastic resonant excitation of a bulk acoustic mode by the pump wave, leading to the formation  of a backwards-propagating optical Stokes wave. A key issue with SBS in technologically relevant materials platforms, such as silicon, is   intrinsically low SBS gain, which motivates considerable interest in novel designs for its enhancement (i.e., \cite{van2015interaction}).   In recent years,   composite materials  have been explored theoretically as a means of   controlling   SBS       \cite{smith2015electrostriction,sun2015analytic,smith2016control,smith2016stimulated,su2017theoretical}.  
  
\begin{figure}[t!]
	\centering
		\includegraphics[width=0.99\linewidth]{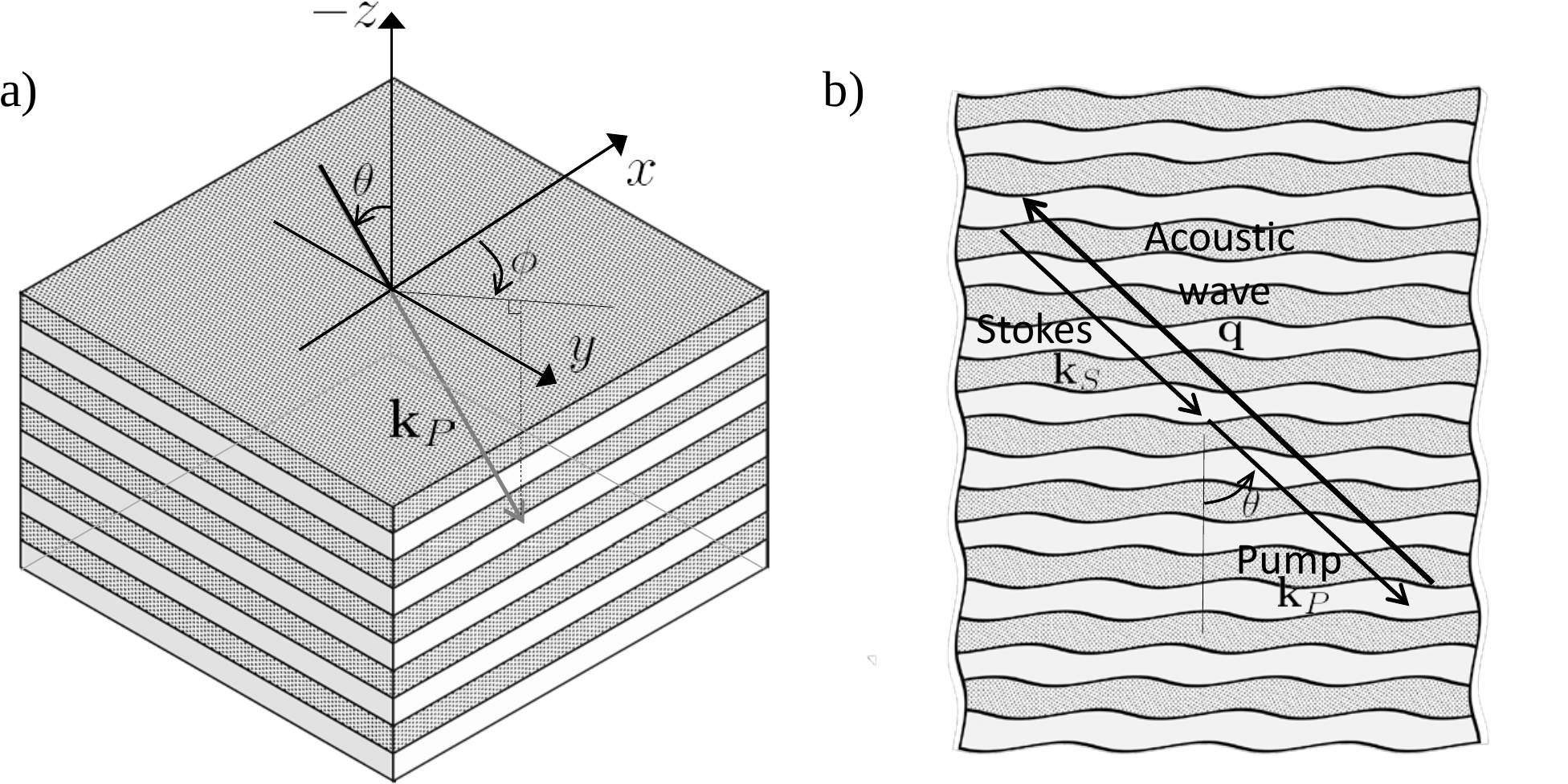}
	\caption{(a) Illustration of composite bulk medium comprising As$_2$S$_3$ glass (white) and silicon (grey) layers   with 	(b) cross-section view showing pump $\bfk_\rmP$, Stokes $\bfk_\rmS$, and acoustic $\bfq$ wave vectors involved in  bulk backwards SBS interaction.
 	 \label{fig:schem}}
\end{figure}

In this work,  we present results for SBS in a   layered medium, as shown in Fig.~\ref{fig:schem}, where in addition to photoelastic processes   (including artificial photoelasticity \cite{smith2017enhanced}, see below), we also have contributions to the SBS gain from  induced optical anisotropy. By incorporating all relevant optoacoustic processes, we show   that the     gain coefficient   of layered media takes values well  above the SBS gain of the constituents in bulk.   To demonstrate this point,  we present results for a   Si--\chalc   layered medium, where we find a gain coefficient of $g_\rmP = 1.28 \times 10^{-9} \, \rmW^{-1} \, \rmm$. Our gain value is two orders of magnitude larger than that of pure Si<100> ($g_\rmP = 2.4 \times 10^{-12}  \, \rmW^{-1} \, \rmm$ \cite{smith2016control}),  75\% larger than that of pure As$_2$S$_3$ ($g_\rmP  = 7.4 \times 10^{-10} \, \rmW^{-1} \, \rmm$ \cite{pant2011chip}), and an order of magnitude larger than results for a suspension  of \chalc   spheres         in Si  ($g_\rmP = 1.06 \times 10^{-10} \, \rmW^{-1} \, \rmm$   \cite{smith2016stimulated}). To the authors' knowledge this   is the first study of SBS in   layered media.   
  
Such enhancements are achieved  due to  changes in the permittivity under     changes in filling fraction      ({\it artificial photoelastic} contributions), as well as  changes to the permittivity under infinitesimal rotation    ({\it roto-optic} contributions \cite{toupin1956elastic,nelson1970measurement,nelson1970measurementerr,nelson1979electric,vacher2006Brillouin}), in addition to {\it intrinsic} photoelastic contributions from   the layers \cite{smith2017enhanced}. 
 
As an optoacoustic process, roto-opticity is not new \cite{toupin1956elastic}, but is much less well-known than conventional photoelasticity, as rotationally induced birefringence is not observed in   materials of high symmetry. Artificial photoelastic effects are much more recent and were discovered by   the authors  in 2015   \cite{smith2015electrostriction}.   All microscopic-scale interactions are implicitly captured in the effective medium treatment, and are contained within the artificial electrostriction contributions (see discussion in Smith {\it et al.}~\cite{smith2017enhanced}). The nontrivial relationship between radiation pressure and artificial electrostrictrion is demonstrated by the fact that artificial contributions vanish in the absence of either a permittivity or stiffness contrast \cite{smith2017enhanced} , whereas radiation pressure effects vanish in the absence of only a permittivity contrast \cite{wolff2014stimulated}. In materials possessing optical isotropy and acoustic anisotropy, such as Germanium,  a material frame  rotation has been  shown to improve the  confinement of   longitudinal acoustic modes   for   SBS \cite{wolff2014germanium}.  

We begin by presenting a generalisation of the coupled-mode approach in Wolff {\it et~al.} \cite{wolff2014stimulated} for   evaluating the SBS gain  of uniform anisotropic media. Assuming  linear constitutive behaviour in both the optical and acoustic properties, the  SBS   gain coefficient (in the absence of irreversible forces) is defined as \cite{wolff2014stimulated}
 \begin{equation}
 \label{eq:GeneralPractice}
 g_\rmP =   \frac{2\omega \Omega  \gamma^2 }{\mathcal{I}_{\rmP} \mathcal{I}_{\rmS}  \mathcal{I}_\rmA \alpha },
 \end{equation}
 where $\gamma$ (units $\left[\mathrm{Pa}\right]$) denotes the total electrostrictive coefficient, $\mathcal{I}_\rmP$ is the pump  intensity (units $\left[\mathrm{W}/\mathrm{m}^2\right]$), $\mathcal{I}_\rmS$ is the Stokes intensity (units $\left[\mathrm{W}/\mathrm{m}^2\right]$), and $\mathcal{I}_\rmA$ is the acoustic intensity (units $\left[\mathrm{W}/\mathrm{m}^2\right]$), where all intensities are associated with the modes of the optical and acoustic waves in the material. Here, $\alpha$ is the   acoustic attenuation  constant   (units $\left[1/\mathrm{m}\right]$).  These   are   given by
 \begin{subequations}
\label{eq:alltheoverlaps}
\begin{align}  
\gamma &= -\varepsilon_0 \, \varepsilon_{im} \, \varepsilon_{jn}  \, P_{mnkl}  \,  \left( \partial_l u_k \right)^\ast \, E_i^\rmP \,  ( E_j^\rmS  )^\ast, \\
\mathcal{I}_{\rmP} &= 2 \,  (\widehat{v}_\rmg^\rmP  )_i \, \epsilon_{ijk}  (E_j^\rmP  )^\ast H_k^\rmP,\\
 \mathcal{I}_{\rmS} &= 2 \,  ( \widehat{v}_\rmg^\rmP  )_i \, \epsilon_{ijk}  (E_j^\rmS  )^\ast H_k^\rmS,\\
 \mathcal{I}_{\rmA} &= -2 \rmi \Omega \,    (u_i)^\ast \, ( \widehat{v}_\rmg^\rmP  )_j  \, C_{ijkl} \, s_{kl},\\
 \alpha &=   {\Omega^2} \mathcal{I}_\rmA^{-1} \, s_{ij} \, \eta_{ijkl} \, (s_{kl})^\ast, 
\end{align}
\end{subequations} 
\noindent  respectively, where $\varepsilon_{ij}$ is the relative permittivity tensor, $u_k$ is the acoustic displacement from equilibrium, $E^{\rmP,\rmS}$ and $H^{\rmP, \rmS}$ are the electric and magnetic field vectors for the pump and Stokes fields, respectively. Additionally, $\widehat{v}_\rmg^\rmP$ is the normalised group velocity for the pump field, $\epsilon_{ijk}$  is the Levi-Civita tensor,  $C_{ijkl}$ is the stiffness tensor, and $\eta_{ijkl}$ is the phonon viscosity tensor. Finally, $\omega$ and $\Omega$ are the angular frequencies of the pump field  and the acoustic wave, respectively, and $s_{kl} = (\partial_k u_l + \partial_l u_k)/2$ is the linear strain tensor.  Here we define $P_{ijkl}$ via  \cite{nelson1970new}
\begin{subequations}
 \begin{equation}
 \Delta \varepsilon_{ij} = -\varepsilon_{im} \varepsilon_{jn} P_{(mn) kl}  \, \partial_l u_k,
 \end{equation}
 where the full photoelastic tensor decomposes as
\begin{equation}
\label{eq:totalphoto}
P_{(ij)kl}  =  
p_{(ij)(kl)}
+ 
p_{(ij)[kl]}  ,
\end{equation}
\end{subequations}
for  non-piezoelectric dielectric materials,  with $p_{(ij)(kl)}$ and $p_{(ij)[kl]}$  denoting the symmetric and antisymmetric (roto-optic) photoelastic tensors, respectively. Following \cite{nelson1970new},  we represent index pair interchange symmetry with parentheses and interchange antisymmetry with square brackets. Note that \eqref{eq:totalphoto} represents a key departure from conventional treatments of SBS, as $p_{(ij)\left[kl\right]}\equiv 0$ in optically isotropic materials \cite{nelson1972brillouin}.

To evaluate \eqref{eq:GeneralPractice}, it is necessary to   determine a large number of modal fields belonging to several different families. For determining the optical fields and quantities in \eqref{eq:alltheoverlaps}, such as  wave polarisations and the refractive index,  we consider   Maxwell's equations with the plane wave ansatz    $E_j = \tilde{E}_j \mathrm{exp}(\rmi k_i x_i  - \rmi \omega t)$ where $\tilde{E}_j$ denotes the polarisation of the wave, and $k_j$ is    the wave vector. This   ansatz admits the system \cite{born1964principles}  
 \begin{equation}
 \label{eq:opteq}
\left( \hat{k}_i \hat{k}_j - \left( 1 - \frac{\varepsilon_j}{n^2}\right) \delta_{ij} \right) \tilde{E_j}  =A_{ij}^{\rmM} \, \tilde{E_j}= 0,
 \end{equation}
 where   $k_i =k \, \hat{k}_i$,   $k  = n \omega / c_0$ is the wave number,     $n$ is the refractive index, and $c_0$ denotes the speed of light in vacuum. Subsequently, for a given wave vector $k_j$, the supported refractive indices   are   obtained by solving $\mathrm{det}(A^{\rmM}) =0$ and the associated eigen-polarisations  $\tilde{E_j}$ are given by the  eigenvectors   of $ A^{\rmM}$.    The corresponding group velocity of the wave ($\left[v_g\right]_i = \partial_{k_{i}} \omega$) is obtained   by implicit differentiation of   \eqref{eq:opteq}. 

  In order to determine the properties of the available Stokes waves for a given pump wave vector $k_j^\rmP$, namely the wave vectors $k_j^\rmS$ and wave polarisations $\tilde{E}_j^\rmS$, we  impose that the direction of the  group velocity vector for the Stokes wave  is opposite  in sign to  that of  the   pump wave ($[ \hat{v}_g^\rmS]_j = - [\hat{v}_g^\rmP]_j$), which is consistent with backwards SBS coupling. From this group velocity condition,  the properties of the Stokes wave  are obtained  following the   procedure above for the pump wave.

\begin{figure*}[t!]
	\centering
		\includegraphics[width=0.42\linewidth]{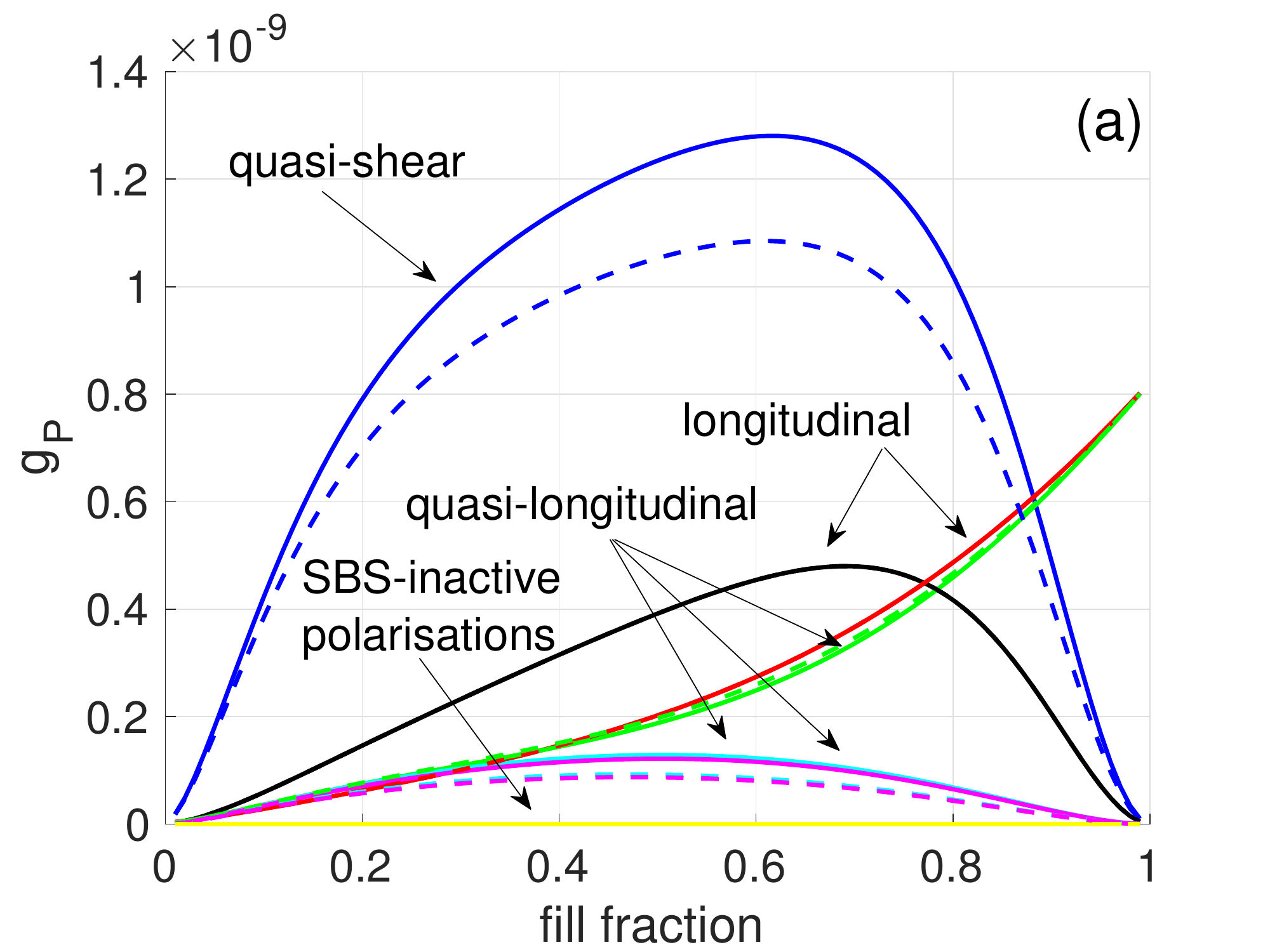}
		\hspace{3mm}
				\includegraphics[width=0.49\linewidth]{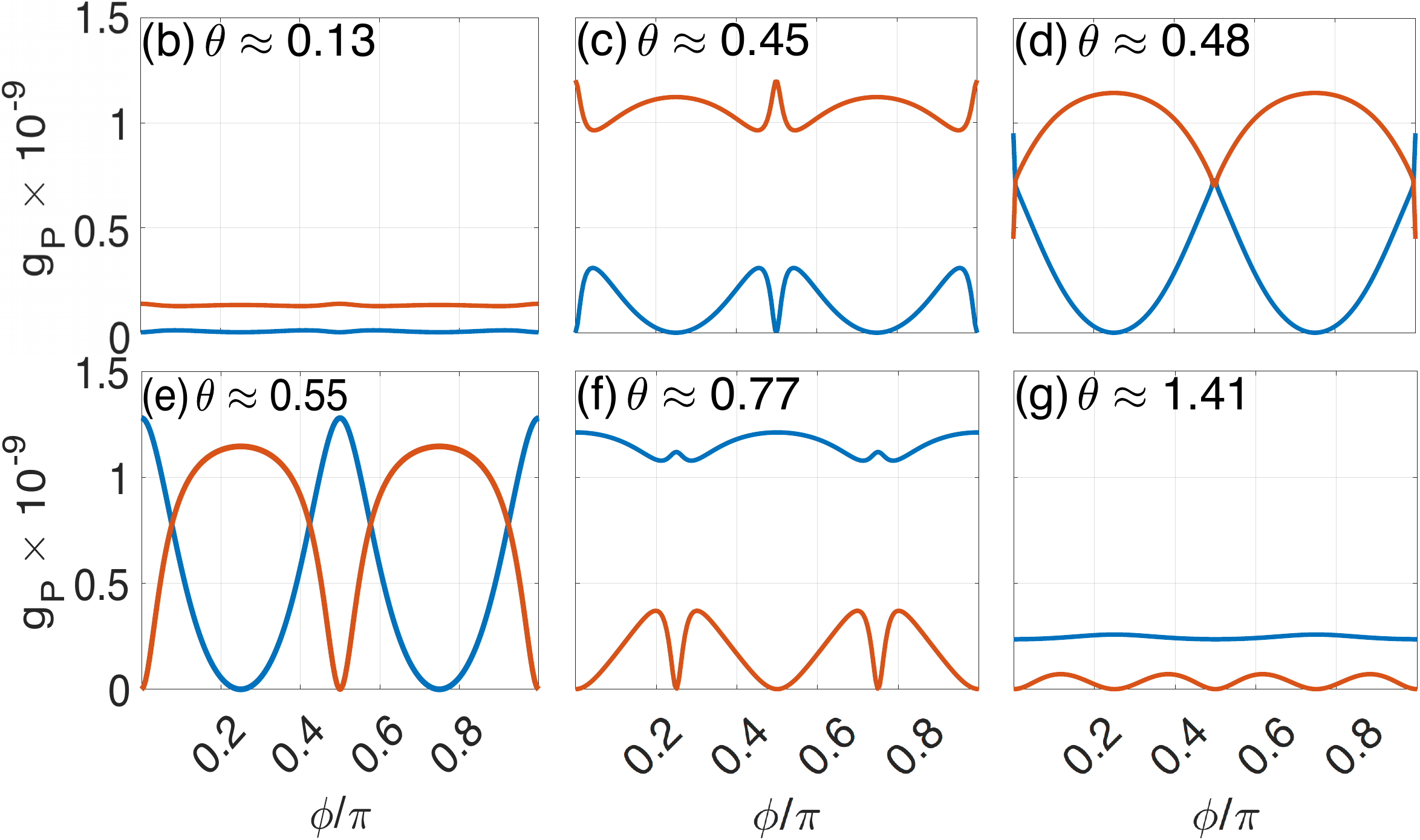}
	\caption{SBS gain coefficient for As$_2$S$_3$-Si layered medium  as      (a)   function of filling fraction $f$ for fixed pump wave vector $k^\rmP \approx (0.52,   0 ,   0.85)$. Dashed curves are gain values when   roto-optic contributions are neglected (i.e., $p_{ij\left[kl\right]}=0$). Each curve represents   different combinations (branches) of pump, Stokes, and acoustic waves (i.e., blue = quasi-shear,     red = longitudinal acoustic wave polarisations); (b-g)    function of   $k^\rmP=( \sin \theta \cos \phi,  \sin \theta \sin\phi ,   \cos \theta)$  at   $f \approx 0.61$. Two  competing mixed  acoustic mode branches  are shown (remaining branches omitted for clarity). In all figures we consider a period of $50$ nm.   \label{fig:as2s3si} }
\end{figure*}

 To determine   acoustic fields and quantities,  we consider the   acoustic wave equation  \cite{auld1973acoustic} assuming the plane wave ansatz
$  u_j = \tilde{u}_j \, \mathrm{exp}({\rmi q_i x_i - \rmi \Omega t})$, where $\tilde{u}_j$ is the polarisation, and the acoustic wave vector $q_j$ is defined      by         $q_j = k_j^\rmP - k_j^\rmS  = q \hat{q}_j$ \cite{boyd2003nonlinear,powers2011fundamentals}. Subsequently,  we    obtain the eigenvalue problem
 \begin{equation}
 \label{eq:christoffel}
 A_{ij}^\rmA v_j= \frac{\rho \Omega^2}{q^2} v_i,
 \end{equation} 
 which returns $\rho \Omega^2/q^2$ as eigenvalues and  $\tilde{u}_j$ as eigenvectors, where    $\rho$ denotes the mass density  and $v_j = \partial_t{u_j}$   the phase velocity of the wave.  The   Christoffel tensor $A_{ij}^\rmA$ is   widely tabulated 
 for a broad selection of  Bravais lattice classes   
 \cite{auld1973acoustic,nye1985physical,newnham2004properties}. 
 
We now proceed to a numerical study examining the SBS performance of Si-\chalc layered media with a unit cell length of $50 \, \rmn \rmm$ and at a vacuum wavelength $\lambda_0 = 1550 \, \rmn \rmm$. Even though the layered media consists of an acoustically isotropic and cubic medium, the   composite medium is optically uniaxial. In Eqs.~(\ref{eq:GeneralPractice})-(\ref{eq:totalphoto}) above,   we use the closed-form expressions for   $\varepsilon_{ij}$, $C_{ijkl}$,  $\eta_{ijkl}$, $\rho$, and $P_{(ij)kl}$ derived in    Smith {\it et  al. }\cite{smith2017enhanced} for laminate materials possessing tetragonal ($4/mmm$) symmetry  (i.e.,  when the constituent layers   possess isotropic or cubic symmetry). These descriptions are valid provided the wavelengths of all optical and acoustic waves are much longer than the periodic length scale of the laminate;  in this regime   the behaviour of the composite material  is accurately described by an effective uniform material. Material  constants for the constituent layers   are given in  Smith {\it et al.~}\cite{smith2016control}.

 For the tetragonal laminates we consider,         two pump waves are generally   supported for wave vectors     oriented away from  directions of high symmetry (one for the extraordinary and another for the ordinary surface). Thus, for a specified $\hat{k}^\rmP$ in a bulk uniaxial crystal, up to four pair combinations of pump and Stokes waves   may contribute to  an SBS process. For each combination of pump and Stokes wave field,    there are up to three   acoustic wave polarisations supported at long wavelengths, giving a total of twelve possible combinations of pump, Stokes, and acoustic waves to participate in an SBS process in a tetragonal ($4/mmm$) material for a given $\hat{k}^\rmP$.  In a  composite material,  the symmetric photoelastic tensor defined in \eqref{eq:totalphoto} may be decomposed further as
\begin{equation}
\label{eq:decomppsymm}
p_{(ij)(kl)} = p_{(ij)(kl)}^\mathrm{pe} + p_{(ij)(kl)}^\mathrm{art},
\end{equation}
representing  some weighted average of constituent photoelastic terms  and   artificial photoelasticity coefficients, respectively \cite{smith2017enhanced}.
 
In Fig.~\ref{fig:as2s3si}(a) we present the gain coefficient as a function of filling fraction for all twelve possible combinations of pump, Stokes, and acoustic wave polarisations (henceforth we  refer  to each of these twelve combinations as {\it branches}) corresponding to $k^\rmP \approx ( 0.52,   0 ,   0.85)$. This particular wave vector corresponds to the maximum possible gain value of $\mathrm{max}(g_\rmP) = 1.28 \times 10^{-9} \, \rmW^{-1} \, \rmm$   for this   material pair, which occurs   at $f\approx0.61$ (note that two other $k^\rmP$ orientations return the same $\mathrm{max}(g_\rmP)$, see below). Superimposed on this figure are the twelve branches for this structure when   roto-optic contributions are neglected, returning a maximum gain value of $g_\rmP = 1.08 \times 10^{-9} \, \rmW^{-1} \, \rmm$  at $f\approx0.61$, and ultimately revealing a roto-optic enhancement of 19\% to the $\mathrm{max}(g_\rmP)$. These dashed curves also allow us to identify the shear contribution to each mode as a function of filling fraction, revealing that the medium-gain branches (depicted by the black, red, and green curves in Fig.~\ref{fig:as2s3si}(a)) are almost purely  longitudinal acoustic modes, for example.  At filling fractions of approximately $88\%$ and higher, the greatest gain value is achieved for a  longitudinally polarised acoustic mode (red curve), which  is the only  acoustic polarisation that is SBS-active in isotropic materials. Recall that pure  Si<100>, corresponding to $f=0\%$, possesses an   intrinsically low SBS performance with  $g_\rmP  = 2.4 \times 10^{-12}$   \cite{smith2016control}. At $f=100\%$ we obtain $g_\rmP=8.02 \times 10^{-10} \, \rmW^{-1} \, \rmm$ for pure As$_2$S$_3$, which is marginally higher than experimental   results   ($g_\rmP  = 7.4 \times 10^{-10} \, \rmW^{-1} \, \rmm$ in As$_2$S$_3$) \cite{pant2011chip}. Note also that  half of the available mode branches are   not SBS-active for any filling fraction.  If we decompose the total electrostrictive coefficient in \eqref{eq:GeneralPractice} as      $\gamma = \gamma^\mathrm{pe} + \gamma^\mathrm{art} + \gamma^\mathrm{ro}$, where  $\gamma^\mathrm{pe}$ refers to   intrinsic photoelastic contributions alone,  $\gamma^\mathrm{art}$ artificial contributions alone, and $\gamma^\mathrm{ro}$ the gain due to roto-optic contributions alone,  we   find   that  these quantities  amount to  74\%, 18\%,  and 8\%  of the overlap for the maximum gain value observed in Fig.~\ref{fig:as2s3si}(a), respectively. The  modest contributions $\gamma^\mathrm{art}$ and $\gamma^\mathrm{ro}$ are responsible for significant increases in the gain  following \eqref{eq:GeneralPractice}, and    highlight  the importance of accurately capturing all optomechanical processes for SBS calculations in composites. To summarise, we find a maximum gain of $ g_\rmP  = 1.28 \times 10^{-9} \, \rmW^{-1} \, \rmm$ at   $k^\rmP \approx ( 0.52,   0 ,   0.85)$ and $f \approx 0.61$, where  $\Omega/(2\pi)=8.4 \, \mathrm{GHz}$, $\alpha = 16665 \, \rmm^{-1}$, $v_\rmg^\rmA = 2353 \, \rmm / \rms$ and $k^\rmS=-k^\rmP$ (materials tensors      are also   summarised in Table \ref{tab:table1}).
   
  In Figs.~\ref{fig:as2s3si}(b-g), we consider the gain coefficient at fixed filling fraction   $f \approx 0.61$ as we examine the     parameter space for the pump wave vector, i.e. for  $k^\rmP=( \sin \theta \cos \phi,  \sin \theta \sin\phi ,   \cos \theta)$, where $\phi$ is measured relative to the $\langle100\rangle$ axis of Si and the layers are stacked along $\langle001\rangle$. Specifically, we present the gain along    $\phi$ (equatorial plane) as we sweep meridional angles $\theta$. In this instance, only the two largest and competing     branches are shown for clarity. These figures clearly demonstrate the   importance of correctly orienting the pump wave vector, as particular orientation angles correspond to gain suppression (i.e., at $\theta \approx 0.45$ and $\phi = \pi/2$), and that the maximum gain value    occurs not only at a single pump wave vector. That is,   the maximum SBS gain is achieved at three pump wavevectors corresponding to $\phi=0,\pi/2,\pi$ with $\theta \approx 0.55$.  For $\theta \approx 0.55$  the maximum gain branch (blue curve) shows that the incident wave vector must be appropriately oriented to achieve maximal results, however, the presence of the second acoustic branch (red curve) ensures that the gain coefficient does not take values beneath $g_\rmP \approx 7.8\times 10^{-10} \, \rmW^{-1} \, \rmm$.   Interestingly, all figures demonstrate the   intense   competition between acoustic mode branches  for the maximum gain position, in addition to  revealing intrinsic symmetries for the gain parameter in layered media as a function of the $k^\rmP$ direction.
  
\begin{figure}[t!]
	\centering
		\includegraphics[width=0.9\linewidth]{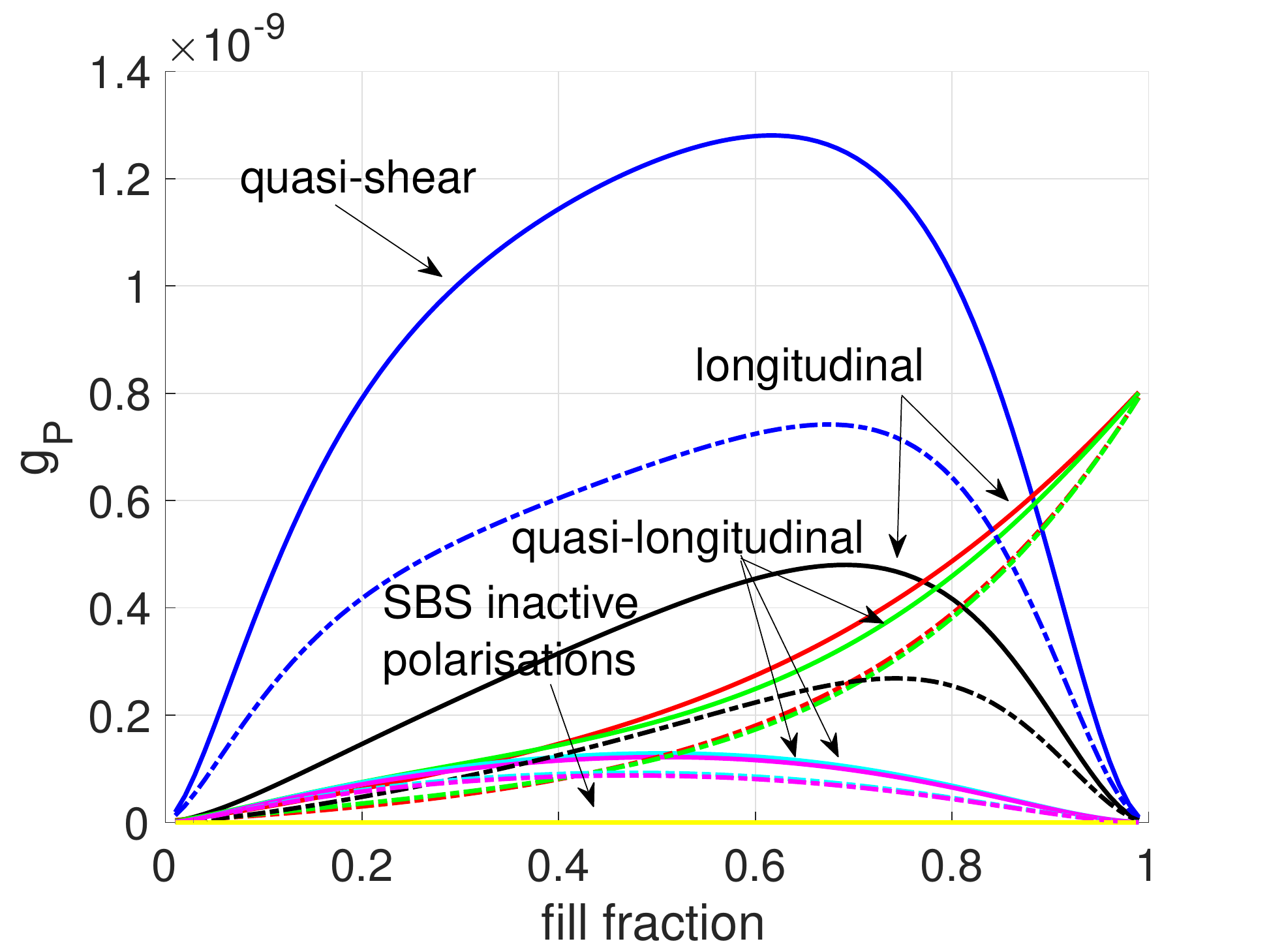}
		\caption{SBS gain coefficient as function of $f$  for As$_2$S$_3$-Si  layered medium shown in Fig.~\ref{fig:as2s3si}(a) where results for neglected roto-optic and artificial photoelastic contributions     (i.e., both $p_{ij\left[kl\right]}=0$ and $p_{ij\left(kl\right)}^\mathrm{art}=0$) are overlaid (dot-dash curves). \label{fig:as2s3sibothoff}}
\end{figure}

   In Fig.~\ref{fig:as2s3sibothoff} we examine the gain coefficient for the layered medium ($f=0.61$) along $k^\rmP \approx (0.52,   0 ,   0.85)$, when both roto-optic and artificial photoelastic contributions are neglected. Also superimposed are the total gain coefficient curves  from Fig.~\ref{fig:as2s3si}(a) for reference. Here, we find that the estimated gain for layered media in the absence of these two contributions is significantly reduced (in particular, the maximum gain   occurs at $f \approx 0.67$ with $g_\rmP = 7.4 \times 10^{-10} \, \rmW^{-1} \, \rmm$). At $f \approx 0.61$ we find $g_\rmP  = 7.3 \times 10^{-10} \, \rmW^{-1} \, \rmm$ and subsequently establish   that   artificial and roto-optic effects increase  the maximum SBS gain coefficient observed in Fig.~\ref{fig:as2s3si}(a) by approximately $75\%$.

In summary, we have presented a theoretical framework for investigating SBS in anisotropic materials, and using this framework  we propose a nanoscale  layered material    with an SBS coefficient that outperforms  leading materials platforms. The structure we propose represents an enhancement in the SBS performance of silicon by more than two orders of magnitude, and is achieved by incorporating contributions from the symmetric photoelastic tensor (containing artificial photoelastic contributions only present in composite media), as well as contributions from the antisymmetric photoelastic tensor (roto-optic tensor), where the latter effect arises in all optically anisotropic media. We have also shown that artificial photoelastic and  roto-optic contributions are non-negligible in optically anisotropic materials, contributing significantly  to the  total gain coefficient for our Si-As$_2$S$_3$ structure.  For this material pair, the gain coefficient of the    layered   medium is larger than that of the  embedded  array medium \cite{smith2016stimulated},    due to  stronger artificial  contributions and the emergence of roto-optic contributions. The framework presented here provides extensive scope for the ongoing development of new materials for future photonics and optomechanics research. 

\begin{table*}[hbp]
 \centering
 \caption{\label{tab:table1}   { \bf \small  
Bulk   values for    relative permittivity   $\varepsilon_{j}$,   stiffness     $C_{ij}$ (units  $\left[ \mathrm{GPa} \right]$), 
phonon viscosity   $\eta_{ij}$  (units  $\left[ \mathrm{mPa} \cdot \mathrm{s} \right]$), density $\rho$ (units  $\left[ \mathrm{kg} \cdot \mathrm{m}^{-3}\right]$), symmetric   photoelastic    $p_{i(j)}$, and roto-optic $p_{i\left[j\right]}$ tensors of layered medium ($f=0.61$).   Subscripts are   in Voigt form and $p_{i(j)} = p_{i(j)}^\mathrm{pe} + p_{i(j)}^\mathrm{art}$.}}
  { \small
 \begin{tabular}{l|cc |  ccc ccc|cccc   cc|c } \hline
  Physical quantity  &$\varepsilon_1$  &$\varepsilon_3$   & $C_{11}$& $C_{33}$ & $C_{13}$  & $C_{12}$ & $C_{44}$ & $C_{66}$  & $\eta_{11}$ & $\eta_{33}$ & $\eta_{13}$ & $\eta_{12}$ & $\eta_{44}$ & $\eta_{66}$ &$\rho$ 
 \\
\hline
    Effective   parameter 	  &8.11  & 7.08   & 68.2 & 28.4 & 9.9 & 21.1 & 9.9  & 34.6 & 2.63  & 2.46 & 2.05  & 2.13   & 0.25 & 0.35 &2864     
\\
 \hline
\end{tabular}
}
\vspace{5mm}
{\small
 \begin{tabular}{l|cc   ccc cc|ccccc|cc   cc } \hline
 Physical quantity   &$p_{1(1)}$  &$p_{3(3)}$   & $p_{1(3)}$   & $p_{1(2)}$ & $p_{3(1)}$ &$p_{4(4)}$   & $p_{6(6)}$   &$p_{1(1)}^\mathrm{art}$  &$p_{3(3)}^\mathrm{art}$   & $p_{1(3)}^\mathrm{art}$   & $p_{1(2)}^\mathrm{art}$ & $p_{3(1)}^\mathrm{art}$    & $p_{4\left[4\right]}$ 
 \\
\hline
  Effective   parameter	  &0.015  & 0.258   & 0.143 & 0.107  & 0.209  & -0.0004   & -0.0422  & 0.012 & 0.03  &0.031    & 0.012 & 0.012      & 0.009 
\\
 \hline
\end{tabular}
}
 \end{table*}

\vspace{3mm}
\noindent
 {\bf \large Funding.} Australian Research Council (ARC) (CE110001018, DP160101691). C.~Wolff acknowledge funding from MULTIPLY  fellowships under the Marie Sk\l{}odowska-Curie COFUND Action (grant agreement No. 713694).

\bibliography{layered_pe}

\end{document}